\documentclass{sf2a-conf2018}
\usepackage{graphicx}
\usepackage{hyperref}
\usepackage[]{natbib}  
\usepackage{epstopdf}

\def\BibTeX{{\rm B\kern-.05em{\sc i\kern-.025em b}\kern-.08em
    T\kern-.1667em\lower.7ex\hbox{E}\kern-.125emX}}
\bibpunct{(}{)}{;}{a}{}{,}  %%%%%%%%%%%%%  A&A bibliography style
\newcommand{\keV}{{\mathrm{keV}}}

\newcommand{\phfluence}{{\mathrm{ph}/\mathrm{cm}^2}}

\begin{document}

\TitreGlobal{SF2A 2018}

%%-----------------------------------------------------------------
%%      the top matter
%%

\title{Detection capability of Ultra-Long Gamma-Ray Bursts with the ECLAIRs telescope aboard the SVOM mission}

\runningtitle{Detection capability of ULGRB with SVOM/ECLAIRs}

\author{N. Dagoneau}\address{CEA Paris-Saclay, DRF/IRFU/Département d’Astrophysique, 91191 Gif-sur-Yvette, France}
\author{S. Schanne$^1$}
\author{A. Gros$^1$}
\author{B. Cordier$^1$}

%% Keep this line, even if the page will be settled afterwards.
%\setcounter{page}{237}

%%-----------------------------------------------------------------

\maketitle

%%-----------------------------------------------------------------
%%        The abstract
%% 
%%  Warning!  within the abstract:
%%  - do not use macros. 
%%  - do not use commands like: \cite, \citet, \citep ... etc.

\begin{abstract}
Ultra-long gamma-ray bursts (ULGRBs) have very atypical durations of more than 2000 seconds. Even if their origins are discussed, the SVOM mission with its soft gamma-ray telescope ECLAIRs could detect ULGRBs and increase the sample of the few which have been detected so far by the Burst Alert Telescope aboard the Neil Gehrels Swift Observatory and some other instruments. 
In this paper, after a short description of the SVOM mission, we present methods developed to clean detector images from non-flat background and known source contributions in the onboard imaging process. 
We present an estimate of the ECLAIRs sensitivity to GRBs of various durations. 
Finally we study the capability of the image-trigger to detect ULGRBs.
\end{abstract}

%% Insert the keywords (to appear in the ADS indexing)
%% Keywords must be separated by a comma

%%-----------------------------------------------------------------

\section{Introduction}
\begin{figure}[ht!]
 \centering
 \includegraphics[width=0.99\textwidth,clip]{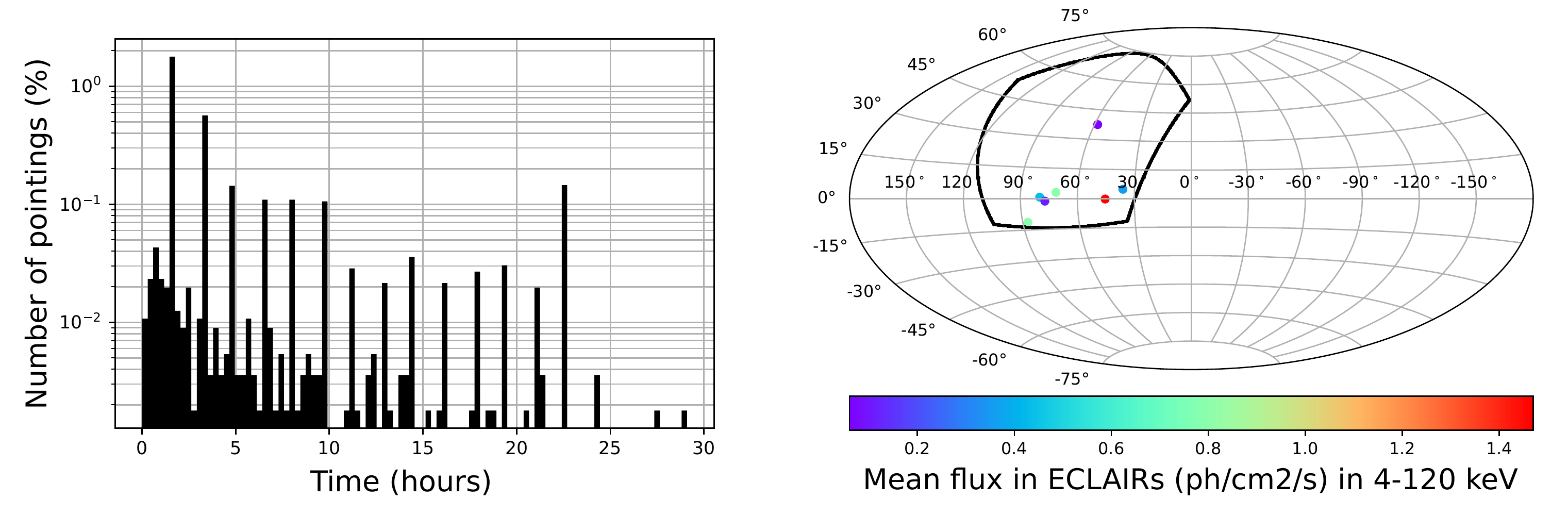}      
  \caption{{\bf Left:} Distribution of the ECLAIRs pointing durations over one year of observations (CNES simulation). The high peaks represent durations which are multiples of a full orbit (about 95 min). Many continuous observations of up to 14 orbits (about 22 h) are foreseen.
  {\bf Right:} Galactic map showing the ECLAIRs field of view (solid black line) of the described pointing to coordinates $(l,b)=(70^o,30^o)$. 7 sources are present in this region of the sky. Color gives the flux in ECLAIRs $4-120~\keV$ band.}
  \label{dagoneau:fig1}
\end{figure}

%%---------------------
It has recently been pointed out that ultra-long gamma-ray bursts (ULGRBs), with very atypical durations of more than 1000 seconds, could form a new class of GRBs \citep{levan_new_2013}. 
ULGRBs could have different progenitors than standard GRBs and be produced by the core collapse of low-metallicity supergiant blue stars \citep{gendre_ultra-long_2013} or the birth of magnetars following the collapse of massive stars \citep{greiner_very_2015}. 
However ULGRBs could also just represent the tail of the standard long GRB distribution \citep{virgili_grb091024a_2013}. 
In any case, it is clear that the duration of these bursts make them peculiar. 
To progress, the sample of the few ULGRBs detected so far by the Burst Alert Telescope onboard the Neil Gehrels Swift Observatory and some other instruments has to be increased. 
This task could be fulfilled by SVOM (Space-based multi-band astronomical Variable Objects Monitor, \citealt{wei_deep_2016}), a French-Chinese mission dedicated to GRBs and transient events, currently under development and scheduled for launch in 2021. 
Thanks to its orbit and pointing strategy  (Fig.~\ref{dagoneau:fig1}, left), the onboard coded-mask telescope ECLAIRs will be able to observe the same portion of the sky continuously during nearly one day. 
With the help of the ECLAIRs image-trigger \citep{schanne2008ICRC....3.1147S, schanne2014arXiv1411.7810S}, searching for long and faint new sources, we expect to increase the sample of ULGRBs.
 
\section{Onboard detection of long duration GRBs with ECLAIRs}
%%-------------------------
\subsection{Long duration imaging process}
%%---------------------

The ECLAIRs coded-mask telescope detects counts, resulting from sky background and source photons which pass through mask holes, as well as internal background. They are accumulated into detector plane images, called shadowgrams. 
In the ECLAIRs image-trigger dedicated to long-duration GRB detection, shadowgrams are built cyclically every $20.48$ s and deconvolved using the coded-mask pattern to produce sky images, which are then stacked together to build images on time-scales $n=1..7$ of duration $2^{n-1} \times 20.48$ s (up to $\sim$20 min), as described in \citealt{schanne2014arXiv1411.7810S}.
The expected ECLAIRs background is mainly composed of the cosmic X-ray background (CXB) modulated along the orbit by Earth passages through the field of view and known sources emitting in soft gamma-rays, rising and setting over the horizon.
In order to be able to detect faint new sources in the reconstructed sky images, the shadowgrams need to be adequately cleaned prior to deconvolution, i.e. the spatial inhomogeneities of the background in the shadowgrams must be removed (to avoid the appearance of artifacts in the reconstructed skies) and contributions of known strong sources must be removed (in order to avoid spurious peaks, called coding noise).

In this cleaning process we first compute a model of the shadowgram, based on a quadratic 2D shape to reproduce the CXB contribution, to which we add an illumination function for each source present in the field of view (which is built using the mask pattern imprinted on the detector, seen from the source position; only the 5 strongest sources are included due to on-board CPU performance constraints). 
This model, with 6 coefficients for the 2D shape and 1 coefficient to reproduce the flux of each source, is fit to the raw shadowgram and subtracted. 
The resulting cleaned shadowgram is then used for deconvolution. 

\begin{figure}[ht!]
 \centering
 \includegraphics[width=0.95\textwidth,clip]{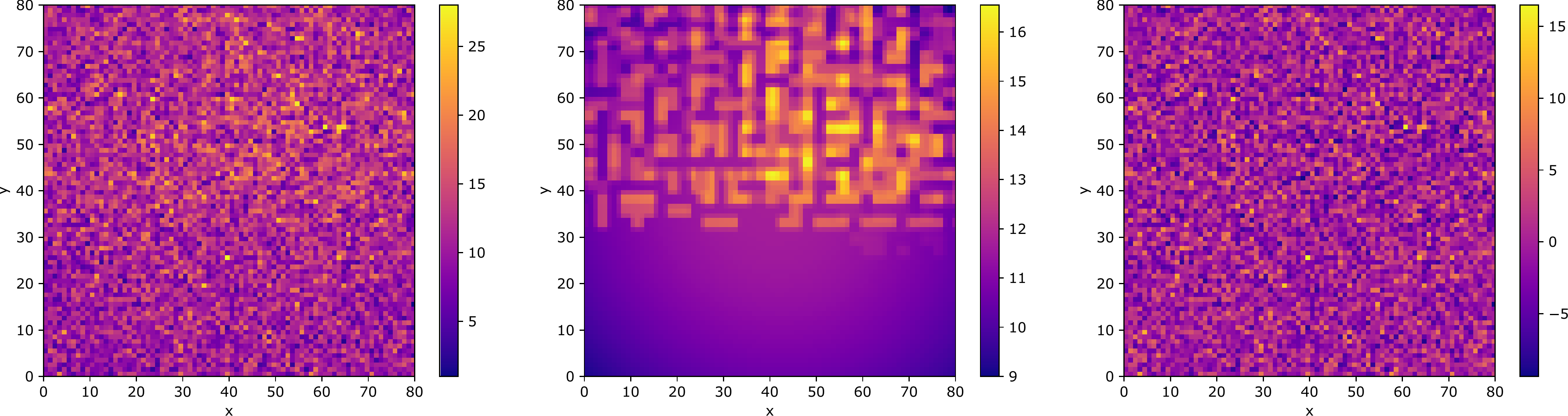}     
  \caption{{\bf Left:} Raw shadowgram in counts. {\bf Center:} Weighted model.  {\bf Right:} Cleaned shadowgram in counts. The $x$ and $y$ coordinates refer to the detector pixels.}
  \label{dagoneau:fig2}
\end{figure}

As an example, a $20.48~\mathrm{s}$ exposure raw shadowgram (the elementary building block of the image-trigger) for a pointing to galactic coordinates $(l,b)=(70^o,30^o)$ is shown in Fig.~\ref{dagoneau:fig2} (left). 
The field of view is shown in Fig.~\ref{dagoneau:fig1} (right) and contains 7 sources: GRS 1915+105, Cyg X-1, Cyg X-2, Cyg X-3, Ser X-1, EXO 2030+375 and Her X-1 (ordered by decreasing flux).

\begin{figure}[ht!]
 \centering
 \includegraphics[width=0.95\textwidth,clip]{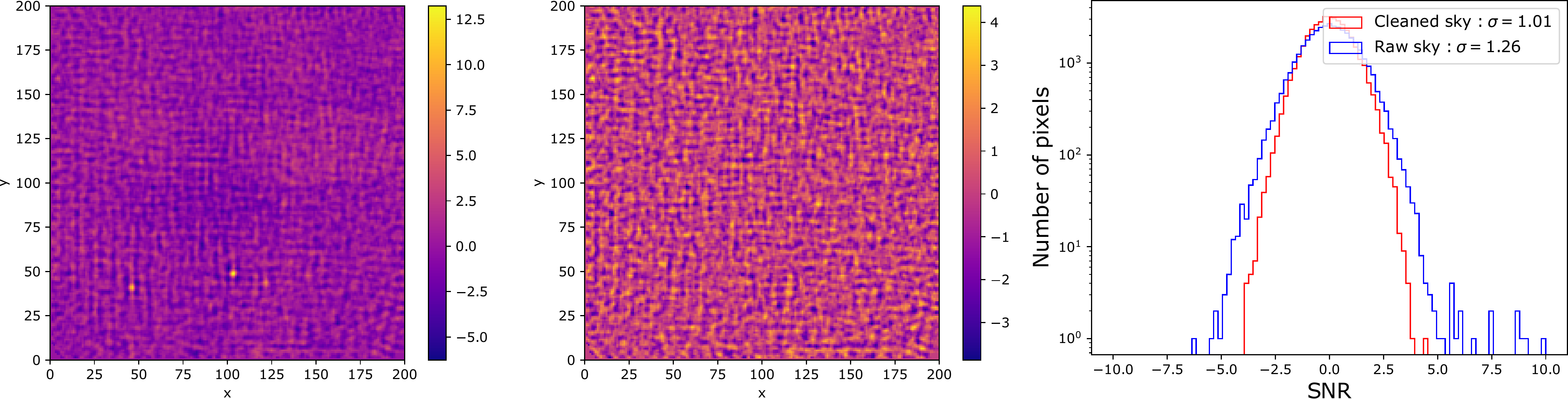}     
  \caption{{\bf Left:} Raw sky in SNR. {\bf Center:} Cleaned sky in SNR.  {\bf Right:} Histogram of sky pixels. The $x$ and $y$ coordinates refer to the reconstructed sky pixels.}
  \label{dagoneau:fig3}
\end{figure}

After subtraction of the model (Fig.~\ref{dagoneau:fig2}, center), we get a cleaned shadowgram (Fig.~\ref{dagoneau:fig2}, right) ready for sky reconstruction. 
For comparison Fig.~\ref{dagoneau:fig3} shows in units of signal-to-noise ratio (SNR) the sky obtained using the raw shadowgram, hereafter raw sky (left), and the sky obtained using the cleaned shadowgram, hereafter cleaned sky (center), as well as the distribution of the SNR of the pixels for both skies (right). 
The distribution is larger for the raw sky (standard deviation $\sigma = 1.26$), which contains residuals from the uncleaned CXB and coding noise from uncleaned sources projected into this reconstructed sky, compared to the cleaned sky ($\sigma = 1.01$).
Furthermore when $20.48~\mathrm{s}$ sky images are stacked together up to the longest time-scale $n=7$ of $\sim20~\mathrm{min}$ considered by the image-trigger, the width of the sky SNR distribution increases significantly for raw skies ($\sigma = 6.19$ after stacking) while remaining almost unchanged for cleaned skies ($\sigma = 1.15$); note that in this example no Earth transits through the field of view are considered.
The threshold for detecting a new source in the trigger is given in units of reconstructed sky pixel SNR, and must be large enough to avoid false triggers. 
With a SNR threshold of $6.5~\sigma$, using cleaned skies the false trigger rate reached is well below 1 per day.
Hence this cleaning process permits detection of faint long-duration new sources.

\subsection{ECLAIRs sensitivity for long duration exposures}
\label{subsec:dagoneau_sensitivity}
\begin{figure}[ht!]
 \centering
 \includegraphics[width=0.5\textwidth,clip]{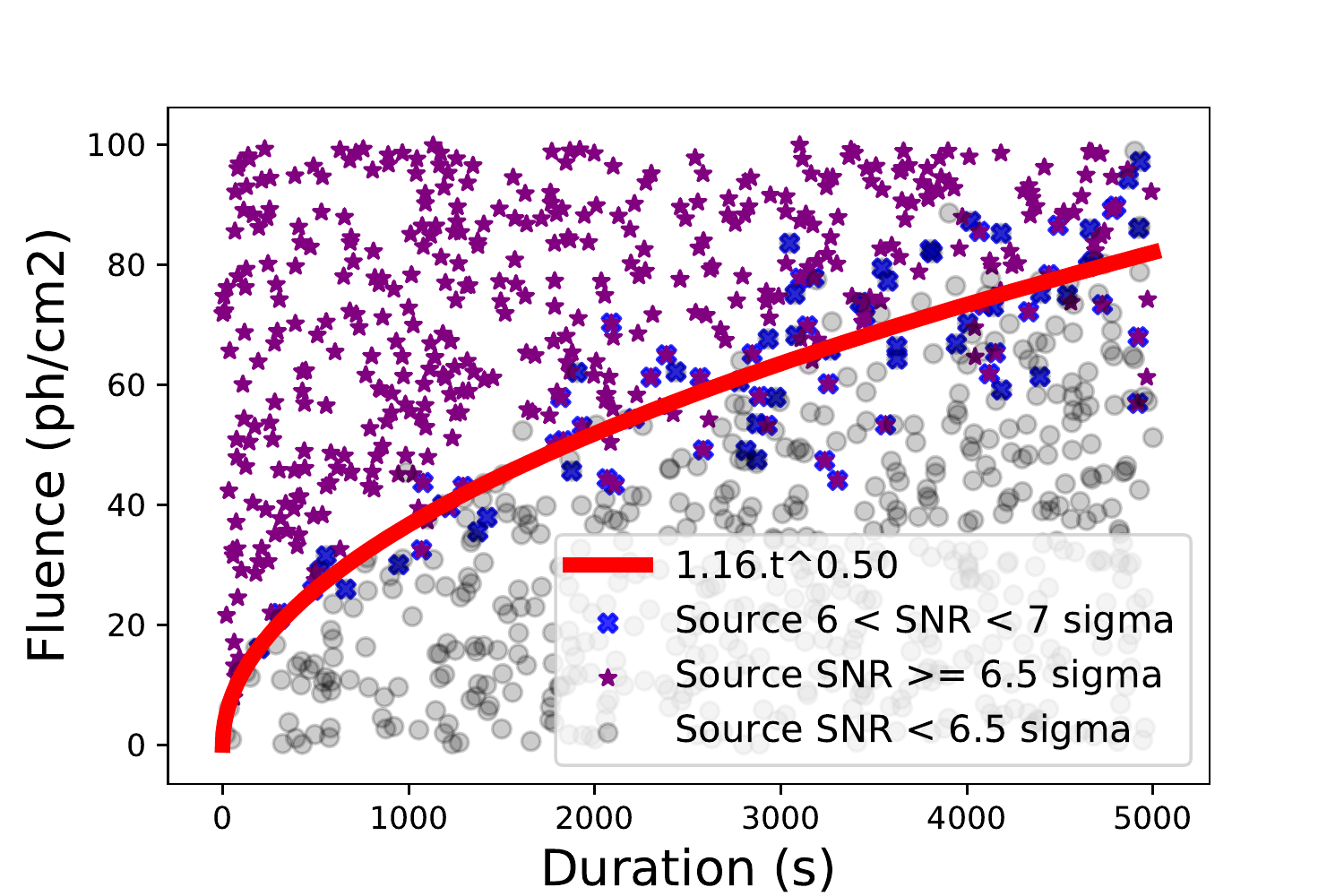}      
  \caption{Fluence threshold in $4-120~\keV$ for canonical GRBs in center on the field of view. Purple stars are events with a SNR in sky image $> 6.5 \sigma$ (hence detected), grey circles are events with a SNR in sky image $< 6.5~\sigma$ (hence not detected) and blue crosses are events with $6 < \mathrm{SNR} < 7~\sigma$ used to fit the square root law.}
  \label{dagoneau:fig4}
\end{figure}

%%---------------------
The ECLAIRs sensitivity to long duration sources is determined for a canonical GRB (with Band spectrum: $\alpha = -1$, $\beta = -2$, $E_{\mathrm{peak}} = 20~\keV$) in the center of the field of view without Earth present. 
We used as background components the CXB and a constant homogeneous internal background on the detector of 0.003 counts/cm$^2$/s/keV. 
10000 simulations for random GRB fluences (between $0.001$ and $100~\phfluence$) and exposure times (between $0.01$ and $5000~\mathrm{s}$) are performed, where each detector image is cleaned (as explained above) and deconvolved.
To evaluate the sensitivity at the SNR threshold set to $6.5~\sigma$, we select GRB-source events from the simulation with a reconstructed SNR at the correct source position such that $6 < \mathrm{SNR} < 7~\sigma$.
In the plot of the fluence as a function of the exposure time (Fig.~\ref{dagoneau:fig4}), these events (blue crosses on the plot) are fit with a square-root law. 
We obtain the following fluence threshold as a function of the exposure time $t$: $f_\mathrm{th} = k\sqrt{t}$ in units of $\phfluence$, with a fit value $k = 1.16~\mathrm{ph}/\mathrm{cm}^2/\mathrm{s}^{1/2}$. 
An on-axis source active during $20$ min should have a typical fluence of about $40~\phfluence$ to be detected by ECLAIRs during that exposure.
Note that this sensitivity curve represents a typical lower bound on the minimal fluence required for source detectability.
Our simulation considered only on axis sources in the center of the field of view, the fluence required will be higher for off-axis sources.
Also the presence of the Earth in a part of the field of view will actually reduce the CXB component of the background on the detector, hence increasing the sensitivity for sources in the sky part not obscured by the Earth.
Additionally the sensitivity curve also depends on the source spectrum.

\section{Towards ultra-long GRBs detection}
%%-------------------------
\subsection{Ultra-long GRBs sample}
%%---------------------
We built a sample of ultra-long GRBs from published literature and from the Swift/BAT GRB-catalog summary tables\footnote{\url{https://swift.gsfc.nasa.gov/results/batgrbcat/summary_cflux/summary_GRBlist/list_ultra_long_GRB_comment.txt}}. 
There are only $21$ ULGRBs known up to now, among which $17$ with BAT data.
We noticed that the available light-curves don't show bright peaks, and although those GRBs are classified as ultra-long, they aren't in fact very long, such that the full durations of those events are difficult to evaluate. 
Nevertheless, Swift has shown through X-ray observations with the XRT that the central engine is active for a longer time than $T_{90}$ measured by BAT. \citealt{zhang_how_2014} and \citealt{boer_are_2015} propose different definitions of the activity time (respectively $T_\mathrm{burst}$ and $T_\mathrm{x}$), with values given in Tab.~\ref{dagoneau:tab1} for the 3 main ultra-long GRBs.

\begin{table}[ht!]
\centering
\begin{tabular}{|l|l|l|l|l|}
\hline
Name       & $T_{90}$ (s)  & $T_{90}$ refined (s) & $T_\mathrm{x} $ (s)  & $T_\mathrm{burst}$ (s)  \\ \hline
GRB111209A & $810.97$ & $13000$       & $25400$ & $63095.73$ \\ \hline
GRB101225A &      & $7000$        & $5296$  & $100000$  \\ \hline
GRB121027A & $80.09$  & $6000$        & $67.38$ & $31622.78$ \\ \hline
\end{tabular}
\caption{Durations of the 3 most cited ULGRBs. $T_{90}$ is from Swift raw data, $T_{90}$ refined is from \citealt{levan_new_2013}, $T_\mathrm{x}$ is from \citealt{boer_are_2015} and $T_\mathrm{burst}$ from \citealt{zhang_how_2014}.}
\label{dagoneau:tab1}
\end{table}

The so-called "ultra long" bursts studied in the literature do not systematically have ultra-long durations in the gamma-ray domain, but have very long activity durations, since the X-ray flashes and internal plateaus are interpreted as manifestations of central engine activity.

Moreover, most ultra-long GRBs have a spectrum available only for part of the light-curve. 
Under these circumstances, it is very difficult to properly simulate the detection by ECLAIRs of the currently known sample of ULGRBs.
For this reason, we decided to study classical long GRBs, and to transport their spectra and light-curves to higher redshifts, in order to create artificially ultra-long GRBs.
We study the detectability of both classical and redshifted long GRBs by propagating them trough the ECLAIRs simulation and by analyzing them with the image-trigger prototype.

\subsection{Classical long GRBs}
\label{subsec:dagoneau_classical_long_grb}
\begin{figure}[ht!]
 \centering
 \includegraphics[width=0.8\textwidth,clip]{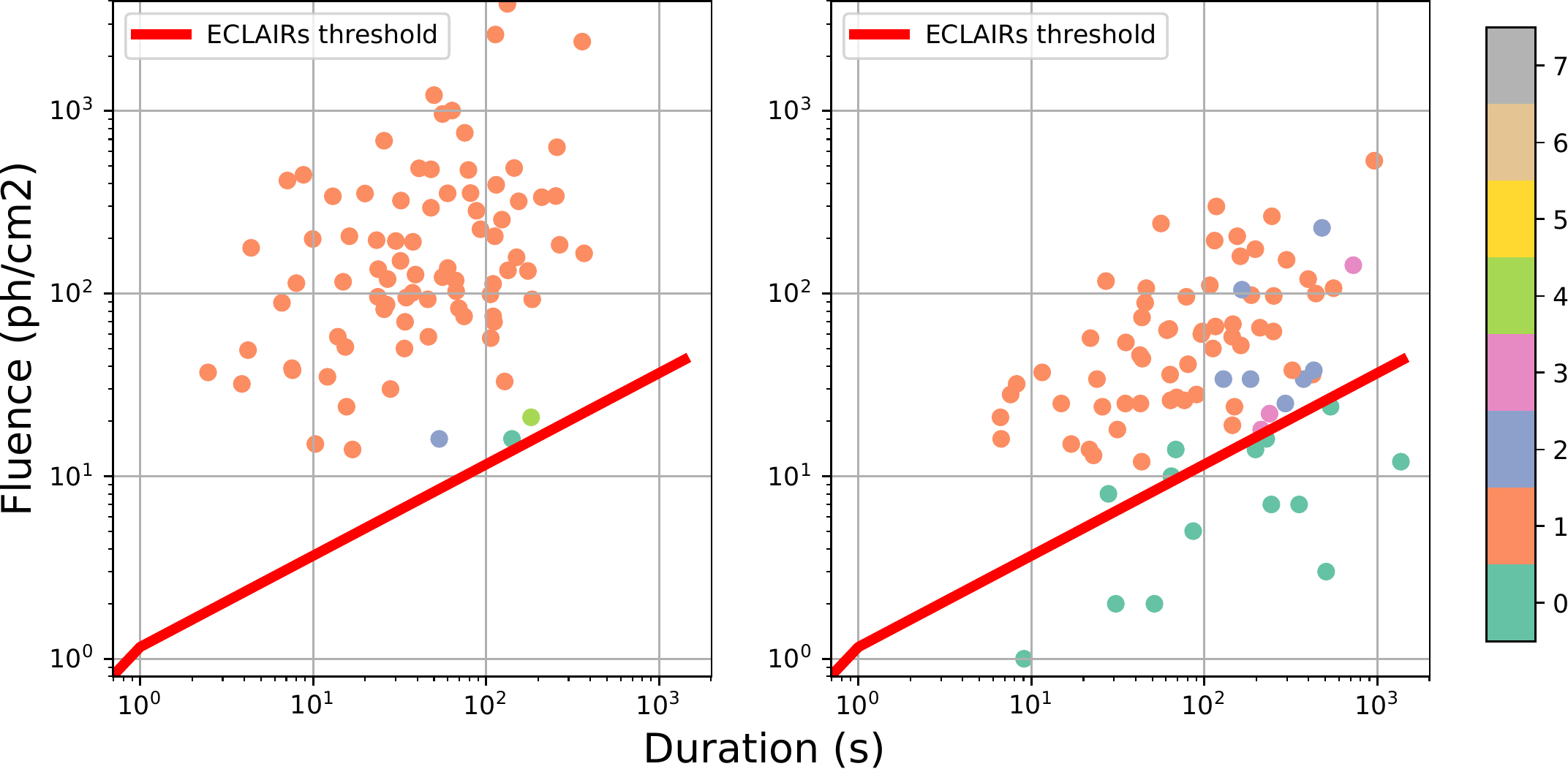}      
  \caption{Fluence-duration plots for GRBs from \cite{heussaff_etude_2015}. Red solid line corresponds to the ECLAIRs on-axis sensitivity limit determined in \ref{subsec:dagoneau_sensitivity}. Colors corresponds to the first time-scale $n$ in which bursts are detected ($n=0$ means no detection). {\bf Left:} At catalog redshift. {\bf Right:} Moved to redshift $z=5$.}
  \label{dagoneau:fig8}
\end{figure}

%%---------------------
We first consider the case of classical long GRBs using the database of GRBs from \cite{heussaff_etude_2015}, which contains 84 GRBs detected by Swift/BAT with known redshift, and simultaneously one other instrument capable of determining a large band spectrum: Wind/Konus in the energy band $10$ - $10000$ keV or Fermi/GBM in $8$ - $1000$ keV. 
Each of these GRBs is placed at the center of the ECLAIRs field of view, extrapolated in the ECLAIRs energy band, and propagated through the ECLAIRs simulator (CxgSim) and the image-trigger prototype software \citep{schanne2014arXiv1411.7810S}.
The distribution of those bursts in the fluence-duration plane is shown in Fig.~\ref{dagoneau:fig8} (left, leaving those bursts at their measured redshift); they are all, except one, above the ECLAIRs on-axis sensitivity.

\begin{figure}[ht!]
 \centering
 \includegraphics[width=0.48\textwidth,clip]{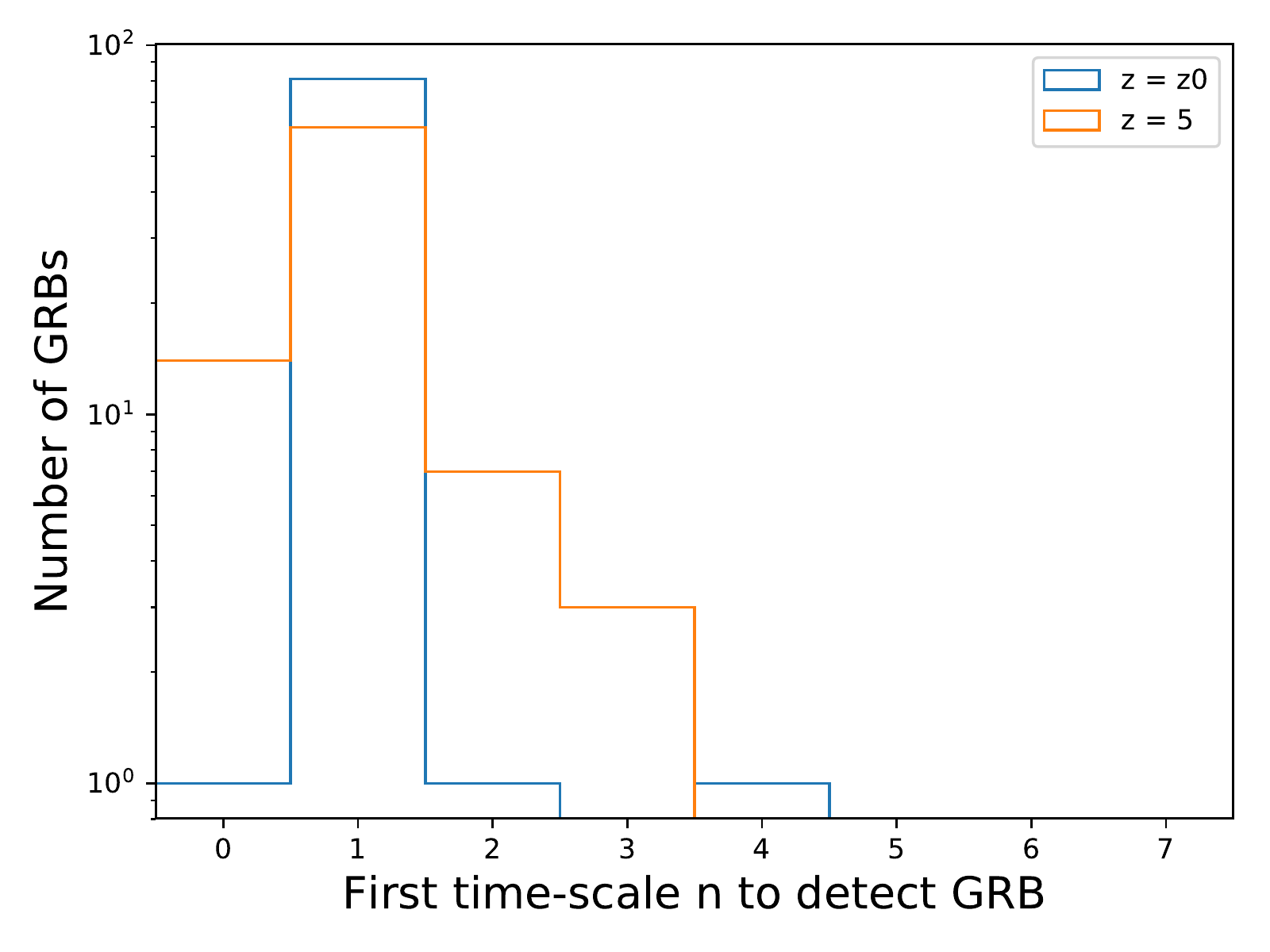}%      
 \includegraphics[width=0.48\textwidth,clip]{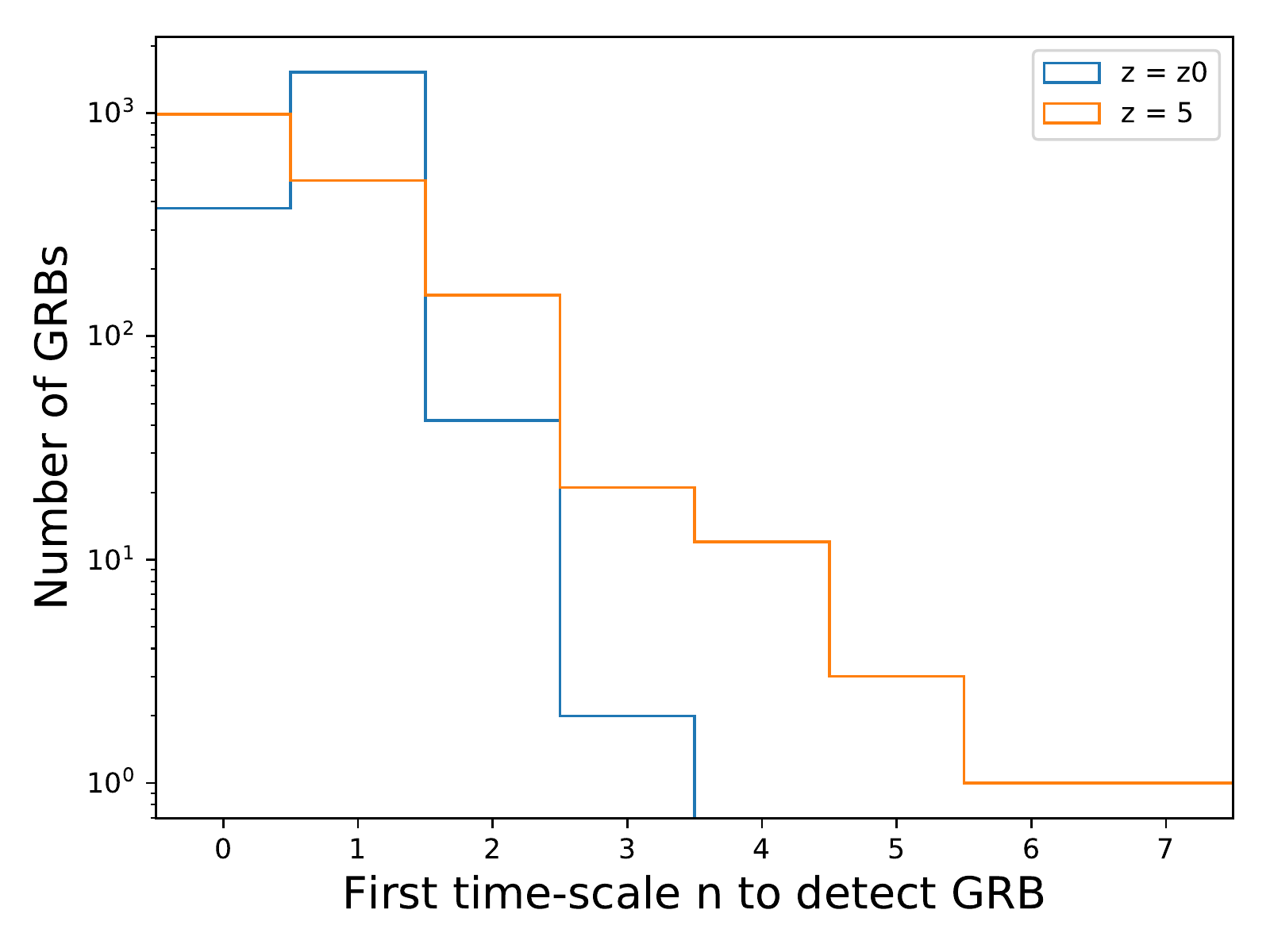}      
  \caption{Histogram of first time-scale $n$ of the image-trigger which detects a GRB from each of the two different databases: {\bf Left:} from \cite{heussaff_etude_2015}. {\bf Right:} from \cite{goldstein_batse_2013}.}
  \label{dagoneau:fig5}
\end{figure}

We are interested in the minimum duration, i.e. the first image-trigger time-scale (indexed $n=1..7$ for durations $2^{n-1} \times 20.48$ s) needed to detect those bursts above the $\mathrm{SNR}>6.5~\sigma$ threshold.
Fig.~\ref{dagoneau:fig5} (left, blue curve) shows the histogram of the first time-scale $n$ in which those bursts are detected.
Among all 84 bursts, one ($<2\%$) is not detected at any time-scale (denoted as $n=0$ in the figure), 81 ($96\%$) are detected within the time-scale $n=1$ (duration $20.48$ s), 1 needs time-scale $n=2$ ($40.96$ s) and a last one needs time-scale $n=4$ ($\sim2.7$ min) to be detected.
We conclude that, out of the the $7$ time-scales analyzed by the image-trigger, no time-scale above $\sim 3$ min is needed to detect GRBs of the \cite{heussaff_etude_2015} database, in the case where they are placed on axis without any other source present in the field of view. This is not surprising since no GRB in the database has a duration exceeding $300$ s.

The same study has been repeated for a larger sample of GRBs, the BATSE database \citep{goldstein_batse_2013}, and the results are shown in Fig.~\ref{dagoneau:fig5} (right, blue curve).
About $78\%$ of the 1940 GRBs of the database are detected in the time-scale $n=1$, while $2\%$ need time-scale $n=3$ or $4$ to be detected, and $\sim 19\%$ are not detected.
The long time-scales, up to $n=7$ implemented in the ECLAIRs image-trigger, are appropriate to detect fainter and/or longer events than those present in the databases used.
In particular such long events could be missing in the BATSE database because BATSE relied on a count-rate trigger only, in which they are difficult to disentangle from background variations.

\subsection{Redshifted long GRBs}
In order to study the detection capability of ultra-long events by the ECLAIRs image-trigger prototype, we used long GRBs from \cite{heussaff_etude_2015} and displace them to higher redshifts, using the program developed by F. Daigne and M. Bocquier (IAP), the procedure being described in \cite{antier-farfar_detection_2016}. 
For each GRB, the list of redshifted photons, with their energy and arrival time, is projected through the ECLAIRs simulator (CxgSim), mixed with background, and processed by the image-trigger prototype software (as in \ref{subsec:dagoneau_classical_long_grb}). 

Using the \cite{heussaff_etude_2015} catalog, and displacing all 84 GRBs to a redshift $z=5$ (Fig. \ref{dagoneau:fig5}, left, orange curve), the number of undetected GRBs rises to 14 ($17\%$) while the number of GRBs detected with time-scale $n=1$ decreases to 60 ($71\%$), the others ($12\%$) are detected at time-scales $n=2$ and $3$, and no longer time-scale is needed in this case neither.
Indeed, redshifting makes GRBs longer but also fainter, eventually placing them below the detection limit, as shown in Fig.~\ref{dagoneau:fig8} (right) where they appear below the sensitivity curve discussed in \ref{subsec:dagoneau_sensitivity}.

Using the larger catalog of \cite{goldstein_batse_2013} with 1940 GRBs (but only 9 have known redshift, for all the others we assume $z=1$), we repeated the procedure to displace them to $z=5$, providing effectively longer lightcurves for 1676 GRBs of the catalog (the others become too faint to be considered).
Fig.~\ref{dagoneau:fig5} (right, orange curve) shows the result: most of the GRBs are still seen in the shortest time-scale ($n=1$), however more than previously need longer time-scales. One burst even needs the longest time-scale $n=7$ to be detected (while it was detected at time-scale $n=1$ before redshift displacement).
It is interesting to note that this burst, GRB970110, was claimed to be a magnetar flare in a nearby ($5.9$ Mpc) galaxy \citep{crider_magnetar_2006}.

\section{Conclusions}
%%--------------------
In the context of the ECLAIRs coded-mask telescope aboard the SVOM mission, this paper presented a procedure developed to remove the CXB and strong known source contributions from the detector images before sky images reconstruction, needed for the imaging of long duration time-scales.
It has been applied to study the detectability of ultra-long GRBs.
However since simulations of existing events of this kind are difficult due to lack of data sets with complete spectral and timing information, we used classical and redshifted long GRBs from known catalogs instead.
One result is that, classical long GRBs do not need very long time-scales in the image-trigger to be detected. 
But once displaced to high redshifts, despites the small number of events, a few bursts need indeed longer time-scales to be detected. 
Such long duration time-scales are not computationally costly in the image-trigger. Given the SVOM pointing law, they should be kept in the flight software, since they provide a space of discovery for ultra-long events, in particular other types of ultra-long transients such as TDEs and SN-shock-breakouts, which remain to be studied in a future work.

% Optional acknowledgements
% -------------------------
\begin{acknowledgements}
ECLAIRs is a cooperation between CNES, CEA and CNRS, with CNES acting as prime contractor. This work is supported by the CEA and by the "IDI 2017" project of the French "Investissements d'Avenir" program, financed by IDEX Paris-Saclay, ANR-11-IDEX-0003-02.
\end{acknowledgements}

%%-----------------------------
%%   Bibliography
%%-----------------------------
%% The following lines are required when using BibTEX (strongly encouraged!):
\bibliographystyle{aa}  % A&A bibliography style file (aa.bst)
\bibliography{dagoneau} % your references in file: Yourfile.bib

\end{document}